\documentclass[twoside]{ilcws10}
\usepackage[latin1]{inputenc}
\usepackage[dvips]{graphicx,epsfig,color}
\usepackage{wrapfig,rotating}
\usepackage{amssymb,amsmath,array}

\begin{document}
\title{KEK GRID for ILC Experiments} 
\author{Akiya Miyamoto, Go Iwai, and Katsumasa Ikematsu
\vspace{.3cm}\\
High Energy Accelerator Research Organization (KEK),\\
1-1 Oho, Tsukuba, Ibaraki, 305-0801 Japan
}

\maketitle

\begin{abstract}
The LCG GRID system is the indispensable infrastructure 
for large scale computing required for ILC experiments.
It had been used extensively for ILD LOI studies and 
its use will be further increased in coming DBD studies.
Experiences during the LOI era and plan towards DBD study 
in KEK are presented.
\end{abstract}

\section{Introduction}
International Linear Collider(ILC) is a global project and 
a good network connection among participant members through Internet
is a crucial infrastructure for the success.
A GRID system is constructed on the Internet and 
it provides not only CPU and storage for large scale computing 
required for ILC experiments, but also sharing of data among ILC experimentalists.
 
Software bases studies in International Large Detector concept (ILD)\cite{ild:2010zzd} has 
utilized LCG GRID\cite{LCGGRID}; It had been used extensively during 
an era of LOI for Monte Carlo production and sharing of produced data 
among members. Especially, there was a strong need in Japan 
to access MC DST samples of about 5 TB placed in Europe.  LCG GRID had 
been used for file transfers successfully.  

Experiences during this period are described in the following sections,
after describing the GRID system in KEK.  A plan towards 
Detector Baseline Design is described subsequently.

\section{Network feature}
A wide-band backbone network has been constructed for HEP community in Japan,
which connects Japanese universities and laboratories participating HEP projects 
such as Belle, J-PARC, ATLAS, ILC, and so on.  In addition, the network covers non-HEP 
users such as material science, bio-chemistry, synchrotron light source and neutron source.
KEK is playing a major role in supporting network services for these activities,
including a GRID deployment and operation. 

The network is connected to the outside of Japan though SINET3\cite{SINET3}.
The SINET3 provides a connection to Hong Kong and Singapore then to 
European network such as GEANT, however the band width to Asian 
countries is limited to 
about less than 1 Gbps. The trans-Pacific network from Japan to US 
has an order wider bandwidth.  Thus the network packet between 
Japan and Europe go through North America, though the actual path
length is longer than a route through Eurasian Continent.

Long distance between Japan and Europe is the limiting factor of
fast transfer of network packets. Typically, we observed a round trip 
time of packets from KEK to IHEP and KISTI, which are institutes in 
China and Korea, to be about 100 msec.  On the other hand, 
those to FNAL in US is about 200 msec and to DESY/IN2P3 in 
Germany/France is about 300 msec.  The round trip time is a 
pedestal time required for every network data transfer, 
independent of a packet size, thus it is not efficient to 
send and receive small files.

\section{GRID system in KEK and experiences in LOI period}
KEK Computing Center is supporting two GRID system, LCG/gLite and RENKEI/NAREGI.
LCG has been used not only by LHC groups but also other HEP groups 
such as Belle, J-PARC and ILC.
RENKEI(REsources liNKage for E-science)\cite{RENKEI} is a research 
to link resources among communities of e-sciences in Japan. It is developing 
NAREGI GRID middle ware. 

For ILC activities, two VOs have been used; CALICE-VO and ILC-VO, which use
LCG GRID middle ware.
CALICE-VO has been used 
by CALICE group\cite{CALICE} for their test beam data analysis and Monte Carlo simulation.
ILC-VO provided CPU resources required for ILD LOI studies.
In Japan, KEK, Kobe university and Tohoku university are joining 
ILC-VO.  During the LOI study period, CPU resources in Japanese GRID sites were very small; It was
less than two order of magnitude smaller than those available at European 
institutes and GRID in Japan had been used mainly for transfers of files 
produced at European institutes or at KEK local batch servers.

For ILD LOI studies, about 70 TB data samples were produced 
mainly at DESY and IN2P3 site.  Sample consisted of simulated, reconstructed
and DST samples produced by processing ILC LOI benchmark
processes\cite{LOIBenchmark} and Standard Model processes at 500 and 250 
GeV center of mass energies. The data size were
placed on GRID SEs for international and inter-regional data accesses.
In the period of LOI studies, 
the time from the production of MC samples to the completion of data 
analysis was limited, thus only DST samples were transferred during the 
period of about 3 moths, except some samples.  The file transfers were 
mainly from Europe to Japan but also partially 
from Japan to Europe. 

In total, about 5 TB data have been transferred
with a typical transfer rate of about 200 kB/sec/port. Due to a limited 
transfer speed, we experienced a frequent time out of transfer, which made it difficult to 
transfer of files with size exceeding 2 GB. The problem had been cured by removing 
the time out limit in file copy.

Large round trip time also imposed a overhead on accesses to the GRID file catalog
located at DESY.  
File sizes of ILD DST files are typically in the order of 10 MB or less. It is 
limited by a CPU time limit of a job for simulation.
In order to transfer many small size file to KEK, 
a tools was developed, which runs at a host of SE and merge many small DSTs
into a compressed single file, then copied it to KEK.

\begin{figure}[htbp]
   \begin{center}
     \includegraphics[width=0.7\columnwidth]{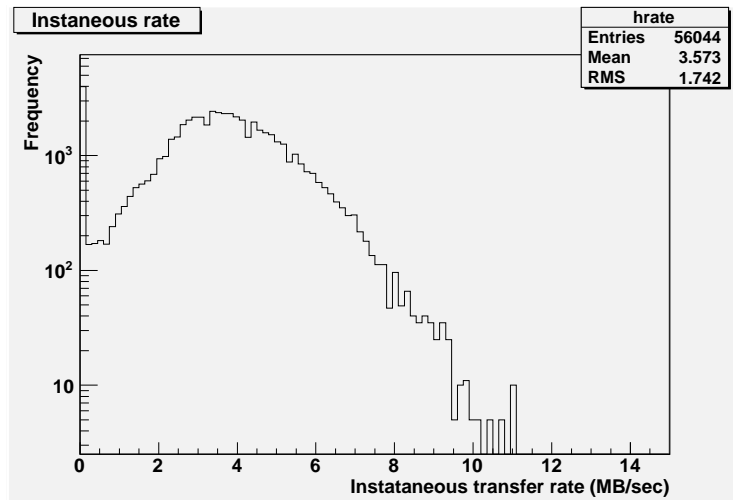}
   \end{center}
   \caption{A typical distribution of instantaneous transfer rate of file transfers 
    from Europe to Japan during the period of ILD LOI studies.}\label{Fig:hrate}
\end{figure}

A typical instantaneous transfer rate during the LOI period were shown in 
Fig.~\ref{Fig:hrate}. The data is a summary of about 3 days transfer using the lcg-cp 
command in December 2008, which is in the middle of the LOI study period.
The most of the entries at 0 transfer rate is the idle time 
before the actual start of the transfer. We used 10 ports for the transfer, thus 
the transfer speed of about 4Mbytes/sec had been achieved.

\begin{figure}[htbp]
   \begin{center}
     \includegraphics[width=0.7\columnwidth]{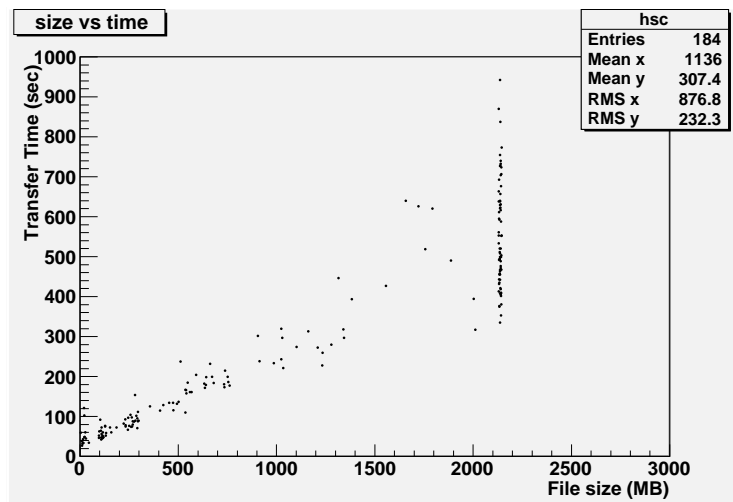}
   \end{center}
   \caption{Scatter plot of file transfer time in sec (vertical axis) and file size in MBytes
( horizontal axis). 
 }\label{Fig:hsc}
\end{figure}
Fig.~\ref{Fig:hsc} is the scatter plot of the transfer time and the file size of 
each file. Each point corresponds to a transfer of each file.  A cluster of points 
at about 2100 MB file size is because of the file size limit set at about 2100 MBytes.
Even with file size close to 0 MB, there are about 20 to 50 seconds pedestal 
in file transfer time, which was mainly caused by the long RTT for access to the catalog information.
From the figure, we see that the file size should be at least a few Mbytes or more 
for efficient file transfer.  Since the speed is limited by RTT, it is important 
to minimize the catalog access at remote site. 

\section{Updates after the LOI period}
CPU resources in KEK were about 0.3 MSI2k (Milion Spec Int2000) during the LOI era.
Thus MC production using KEK GRID resources were limited.  The MC production at KEK site 
had been performed mainly using local batch server systems.
However, it has been extended significantly since the end of the LOI studies.  
There are 5 computing elements (CE)
are operating; about 1600 cores of about 400 CPUs with  about 6M SI2K in total.
We hope that more MC production can be performed using KEK GRID resources in coming 
studies for ILC re-baselining and DBD.

Storage element in KEK has been increased as well. It is using DPM as a SRM.  Backend storage device is IBM HPSS,
TS3500.  In maximum, 3TB data can be stored, which are shared by other VOs and batch server users, 
and actual storage space for ILC VO depends on actual amount of tapes installed.  But, we expect 
the system provides a sufficient storage capacity for coming DBD studies.

\section{Summary}
In summary, 
KEK GRID had been used successfully during the LOI era. Especially, 
GRID played the indispensable role 
for data transfer between Japan and Europe.
In past 12 months, GRID resources in KEK has been increased significantly. 
We hope to be able to contribute significantly in coming MC productions 
towards DBD studies.

\section*{Acknowledgments}.
The author would like to thank KEK Computing Center and GRID teams for supporting 
and helping the use of the GRID system. This work is supported in part by the Creative 
Scientific Research Grants No. 18GS0202 of the Japan Society for Promotion of Science (JSPS),
and Toshiko Yuasa Loboratory (TYL).


\end{document}